\documentclass{vgtc}                          

\ifpdf
  \pdfoutput=1\relax                   
  \usepackage{graphicx}                
  \DeclareGraphicsExtensions{.pdf,.png,.jpg,.jpeg} 
\else
  \usepackage{graphicx}                
  \DeclareGraphicsExtensions{.eps}     
\fi%

\graphicspath{{figures/}{pictures/}{images/}{./}} 

\usepackage{microtype}                 
\PassOptionsToPackage{warn}{textcomp}  
\usepackage{textcomp}                  
\usepackage{mathptmx}                  
\DeclareMathAlphabet{\mathcal}{OMS}{cmsy}{m}{n}
\DeclareSymbolFont{largesymbols}{OMX}{cmex}{m}{n}
\usepackage{multirow}
\usepackage{amsmath}
\usepackage{float}
\usepackage{subfig}
\usepackage{xcolor}
\usepackage{adjustbox}

\usepackage{times}                     
\usepackage{cite}                      
\usepackage{tabu}                      
\usepackage{booktabs}                  

\onlineid{0}

\vgtccategory{Research}

\vgtcinsertpkg

\title{Divide-Conquer-and-Merge: Memory- and Time-Efficient Holographic Displays}

\author{Zhenxing Dong\thanks{Z. Dong and J. Jia are joint first authors and contribute equally.}%
\and Jidong Jia\footnotemark[1] %
\and Yan Li
\and Yuye Ling \thanks{Y. Ling is the corresponding author.\\\indent\indent Email:$\{$d$\_$zhenxing, jjd1123, yan.li, yuye.ling$\}$@sjtu.edu.cn.}}
\affiliation{\scriptsize Department of Electronic Engineering, Shanghai Jiao Tong University}

\teaser{
  \centering
  \includegraphics[width=\linewidth]{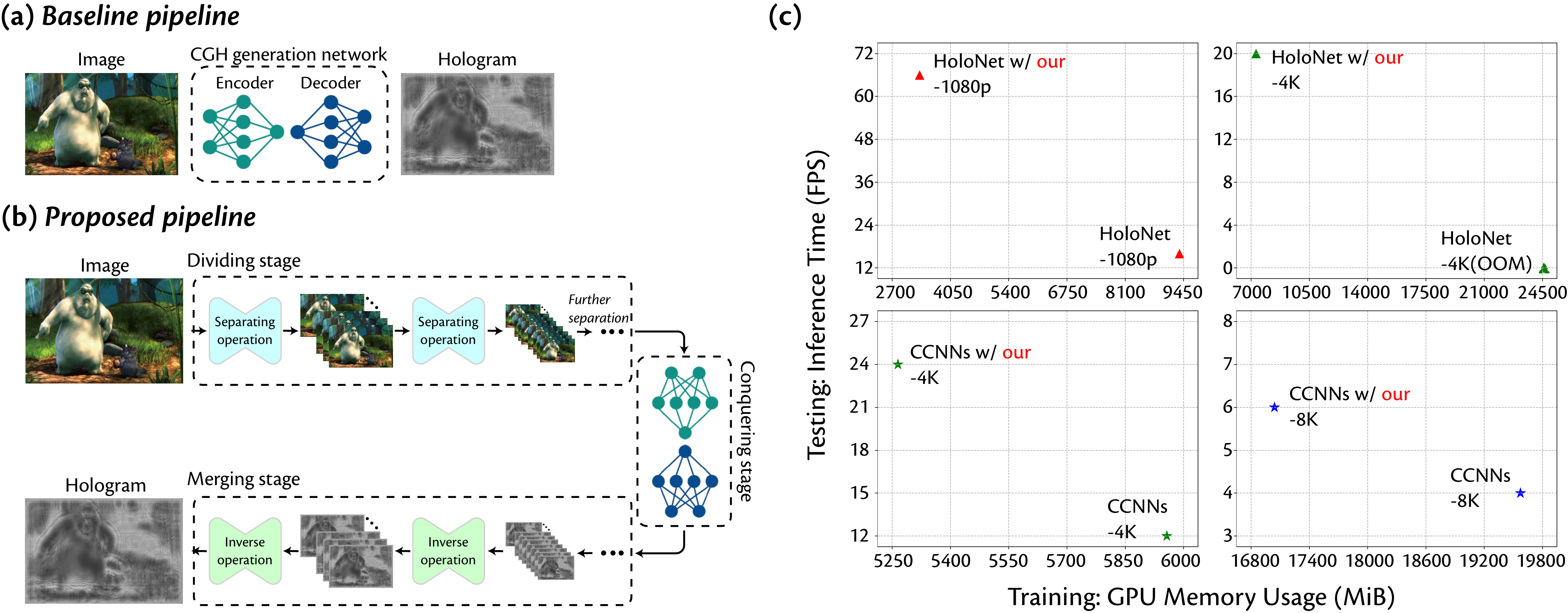}
 \caption{\textbf{Baseline pipeline, Proposed pipeline and Comparison of training GPU memory requirements and testing inference time}. (a) The baseline CGH generation pipeline directly processes the image input within the CGH network to synthesize the phase-only hologram. (b) We propose a divide-conquer-and-merge strategy to enable current holographic pipelines to generate higher-definition holograms at a faster speed. (c) We integrate our strategy into two state-of-the-art CGH generation networks, namely HoloNet and CCNNs, to demonstrate the effectiveness of our strategy. Compared to the baseline pipelines, our strategy meaningfully reduces the memory requirements during the training stage and significantly improves the inference time in the testing stage. The Big Buck Bunny image comes from \href{www.bigbuckbunny.org}{www.bigbuckbunny.org}  (© 2008, Blender Foundation) under the Creative Commons Attribution 3.0 license (\href{https://creativecommons.org/licenses/by/3.0/}{https://creativecommons.org/licenses/by/3.0/}).}
    \label{fig:teaser}
}

\abstract{Recently, deep learning-based computer-generated holography (CGH) has  demonstrated tremendous potential in three-dimensional (3D) displays and yielded impressive display quality. However, most existing deep learning-based CGH techniques can only generate holograms of 1080p resolution, which is far from the ultra-high resolution (16K+) required for practical virtual reality (VR) and augmented reality (AR) applications to support a wide field of view and large eye box. One of the major obstacles in current CGH frameworks lies in the limited memory available on consumer-grade GPUs which could not facilitate the generation of higher-definition holograms. To overcome the aforementioned challenge, we proposed a divide-conquer-and-merge strategy to address the memory and computational capacity scarcity in ultra-high-definition CGH generation. This algorithm empowers existing CGH frameworks to synthesize higher-definition holograms at a faster speed while maintaining high-fidelity image display quality. Both simulations and experiments were conducted to demonstrate the capabilities of the proposed framework. By integrating our strategy into HoloNet and CCNNs, we achieved significant reductions in GPU memory usage during the training period by \textbf{64.3\%} and \textbf{12.9\%}, respectively. Furthermore, we observed substantial speed improvements in hologram generation, with an acceleration of up to 3$\times$ and 2 $\times$, respectively. Particularly, we successfully trained and inferred \textbf{8K} definition holograms on an NVIDIA GeForce RTX 3090 GPU \textit{for the first time} in simulations. Furthermore, we conducted full-color optical experiments to verify the effectiveness of our method. We believe our strategy can provide a novel approach for memory- and time-efficient holographic displays} 

\CCScatlist{
    \CCScatTwelve{Computing Methodologies}{Computer Graphics}{Computer-generated Holography}{}
}

\begin{document}


\firstsection{Introduction}
\maketitle
Holographic displays, utilizing diffractive optical elements for light modulation, enable pixel-level focus control, aberration correction, and visual calibration, which are not achievable with other display technologies \cite{wakunami2016projection}. Therefore, it holds the potential to become the enabling technology for the next-generation virtual and augmented reality (VR/AR) devices \cite{Chang:20}. With advancements in computational algorithms, computer-
generated holography (CGH) has made significant progress in synthesizing holographic patterns through numerical simulation of light propagation. These patterns are loaded onto a spatial light modulator (SLM) \cite{zhang2014fundamentals} in a sequential manner, enabling the dynamic and precise reproduction of virtual object wavefronts.

Deep learning-based CGH methods have delivered exceptionally high-fidelity and photorealistic holographic images without artifacts. Peng et al. \cite{Peng2020} and Chakravarthula et al. \cite{Chakravarthula2020} proposed a novel camera-in-the-loop (CITL) optimization strategy to enhance the display quality of optical experiments. Dong et al. \cite{Dong2023} proposed a Fourier-inspired neural module to improve cross-domain learning in convolutional neural networks (CNNs) to improve the quality of reconstructed images. Choi et al. \cite{Choi2021} introduced a neural network-parameterized multiplane wave propagation model based on CITL, which achieved high-quality 3D holographic images. Shi et al. \cite{Shi2021} presented the lightweight residual CNN in supervised learning and anti-aliasing double phase method (AA-DPM) to synthesize real-time and photorealistic 3D holograms.

Despite these major advancements in display quality, it is still a significant challenge to generate ultra-high-definition holograms. For practical VR and AR applications, ultra-high-definition CGHs (16 K+) are needed to support a wide field of view (FOV) and a large eyebox \cite{Shi:22}. However, the abovementioned holographic networks are mostly limited on generating holograms with a definition of 1080p due to the substantial GPU memory requirements for training epoch. Hence, it is crucial to investigate methodologies for alleviating GPU memory usage during training stage to achieve immersive VR and AR experiences.

Recently, researchers have started exploring new holographic network architectures to generate 4K definition holograms\cite{holoencoder, complex}. Wu et al. \cite{holoencoder} utilized a single auto-encoder network instead of two sub-networks in HoloNet to generate 4K holograms. However, this architecture exhibited a significant display quality degradation. Specifically, compared to HoloNet, the PSNR of its reconstructed images decreased by 6 dB. Zhong et al. \cite{complex} proposed the efficient complex-value neural networks (CCNNs) to learn complex-valued propagation for generating high-quality 4K holograms. While CCNNs achieved a favorable balance between network parameters and reconstructed image quality, they did not further explore the generation of higher-definition holograms.

In this paper, we proposed a novel divide-conquer-and-merge strategy to enable current holographic networks to generate higher-definition holograms, as shown in \autoref{fig:teaser}. Specifically, in the “dividing” stage, our method first applies a pixel-unshuffle layer on the input image to obtain $r^2$ sub-images, where $r$ is the scale factor of the pixel-unshuffle layer. It should be noted that the pixel-unshuffle layer can be replaced by other operations with the same purpose, such as a learnable layer or a CNN. Next, during the “conquering” stage, we predict corresponding lower-definition holograms. Finally, in the “merging” stage, these sub-images are rearranged to form a higher-definition image. To alleviate the degradation in hologram quality, we elaborate a cross-domain SR network that fully integrates the information between neighboring pixels in sub-holograms to generate a large-scale hologram. Furthermore, our framework, like other divide-and-conquer algorithms, can be implemented in a recursive form to further improve the image quality as shown in \autoref{fig:teaser}. We implemented our strategy both numerically and experimentally on two SOTA networks (Holonet and CCNNs) to validate its effectiveness. It turns out that our method can significantly reduce GPU memory usage by \textbf{64.3\%} and \textbf{12.9\%} respectively during the training period, while maintaining the displayed quality of reconstructions. Furthermore, HoloNet with our method achieves the faster generation speed of 66 FPS for 1080p holograms, compared to the na\"ive HoloNet's capability of only 16 FPS. Our approach achieves 2$\times$ and 1.5$\times$ faster generation of 4K and 8K holograms, compared to the baseline CCNNs. Particularly, we successfully trained and inferred \textbf{8K} definition holograms on an NVIDIA GeForce RTX 3090 GPU card \textit{for the first time} in simulations. We believe that our proposed method offers a new perspective on how to resolve the conflict between generating high-definition holograms and the constraints imposed by limited GPU memory. 


Our primary contributions are as follows:
\begin{itemize}
    \item We introduced a divide-conquer-and-merge strategy to address the  memory and computational capacity scarcity in ultra-high-definition CGH generation while ensuring the generation of high-quality phase-only holograms.
    \item We validated the effectiveness of our method through extensive simulations. The simulation results demonstrated that integrating our method into the majority of existing CGH generation frameworks led to smaller GPU memory requirements during the training stage and faster hologram generation speeds in the testing stage.
    \item We developed a full-color holographic display system to conduct optical experiments for a VR demo. Furthermore, we implemented an AR setup to demonstrate the performance of proposed strategy.
\end{itemize}

\section{Related Works}

\subsection{Computer-generated Holography}

CGH is a technique to produce holographic patterns by numerically simulating the light propagation including diffraction and interference. It shows great promise in replicating focus \cite{Choi2021, Shi2021} and parallax cues \cite{Chakravarthula2022}, as well as correcting visual and optical aberrations \cite{wakunami2016projection, 10.1145/3072959.3073624}. CGH utilizes an SLM to reproduce the wavefront from the virtual object. Unfortunately, the current commercial SLMs lack the capability to simultaneously modulate both amplitude and phase, and phase-only SLMs are typically preferred due to their higher diffraction efficiency. Therefore, a meaningful challenge in CGH is how to encode a high-quality phase-only hologram from a complex-valued hologram while ensuring accurate reconstruction of the optical field.

In the past few decades, researchers have developed various CGH algorithms in an attempt to achieve better display quality within limited generation time. These conventional approaches can be broadly divided into two categories: iterative methods \cite{gerchberg1972practical, chakravarthula2019wirtinger, Peng2020} and non-iterative methods \cite{10.1145/3072959.3073624, maimone2020holographic}. Common iterative algorithms, such as Wirtinger holography \cite{chakravarthula2019wirtinger} and stochastic gradient descent (SGD) \cite{Peng2020}, have the potential to achieve high-quality images but are often time-consuming. For instance, SGD can deliver exceptional display quality, but it requires iterative optimization over an extended period to achieve optimal results. While non-iterative algorithms, such as dual-phase amplitude coding (DPAC) \cite{10.1145/3072959.3073624}, provide faster computation compared to iterative methods but can result in images with lower contrast. In addition to the above algorithms, there are also some other CGH approaches \cite{pi2022review} which aim to reduce the computational time and improve the display quality. However, to date, a trade-off between the computational time and the display quality existed among conventional CGH algorithms.

\begin{figure*}[htb]
 \centering 
 \includegraphics[width=\linewidth]{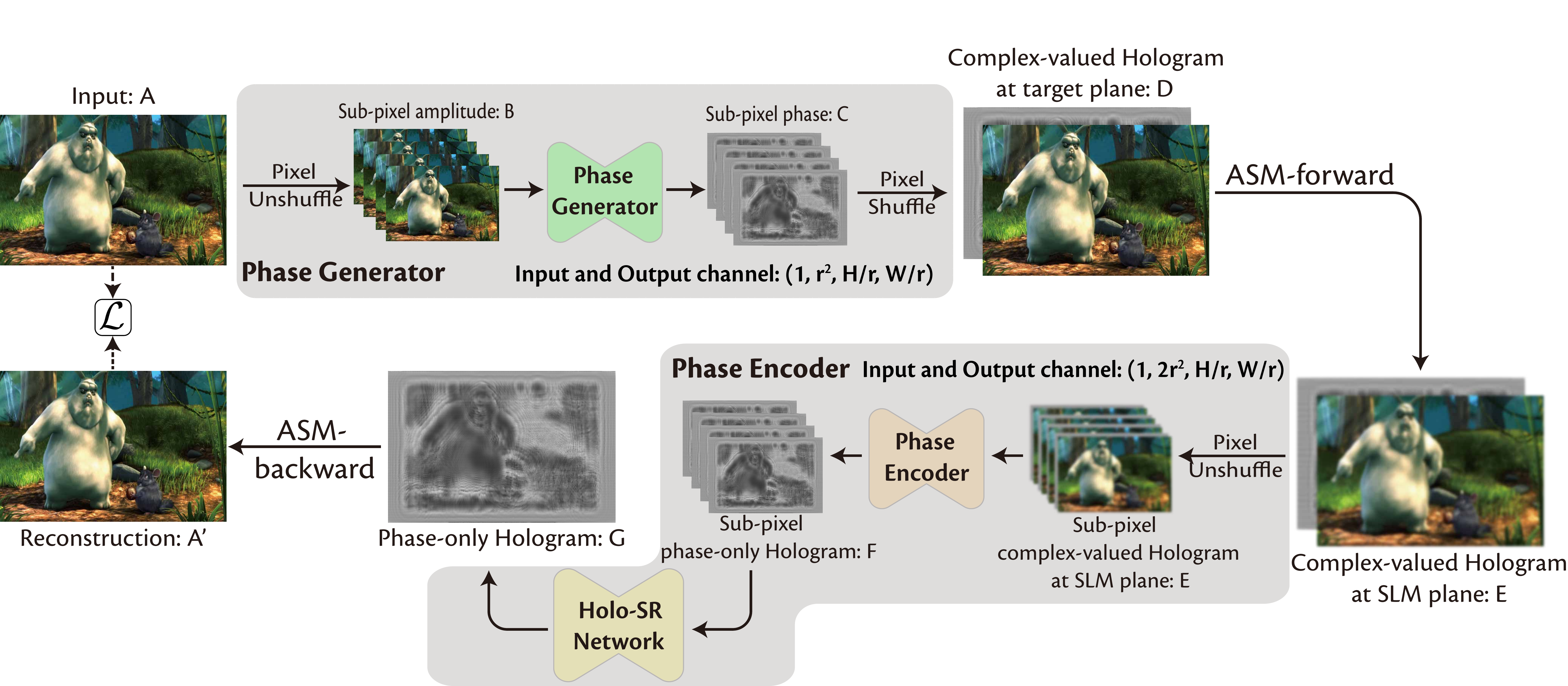}
 \caption{An overview of our proposed framework. The method can be divided into two parts, which are inserted into the phase generator and phase encoder sections of a CGH generation network, respectively. For the phase generator part, the module first performs a pixel-unshuffle operation on an image of size $H\times W$ to get $r^2$ sub-images of size $ H/r\times W/r$. Next, these sub-images are fed into the phase generator of the original network to predict $r^2$ phase sub-images, and then upsamples phase sub-images into a phase image of size $H\times W$ by a pixel-shuffle layer. For the phase encoder part, the module is similar to the phase generator part, moreover for the upsampling step a lightweight SR network is used as a replacement for the pixel-shuffle layer to strengthen the quality of the generated hologram. The Big Buck Bunny image comes from \href{www.bigbuckbunny.org}{www.bigbuckbunny.org}  (© 2008, Blender Foundation) under the Creative Commons Attribution 3.0 license (\href{https://creativecommons.org/licenses/by/3.0/}{https://creativecommons.org/licenses/by/3.0/}).}
 \label{fig:overall_architecture}
\end{figure*}

Recently, deep learning-based approaches have emerged as a promising technique to balance the trade-off between image quality and inference time in CGH. Horisaki et al. \cite{Horisaki} first applied a CNN to infer the hologram from handwritten digital images. Peng et al. \cite{Peng2020} proposed a new CGH architecture, HoloNet, that enabled real-time 2D holographic display with an image quality comparable to that of previous iterative methods. Dong et al. \cite{Dong2023} introduced a neural module inspired by the Fourier transform to enhance cross-domain learning in CNNs and improve the quality of reconstructed images. Choi et al. \cite{Choi2021} further developed the differentiable CITL model and extended its application to multi-plane holography, achieving high-quality 3D displays in optical experiments. Shi et al. \cite{Shi2021} proposed a residual network architecture that is more efficient to synthesize real-time and photorealistic 3D holograms. However, due to the substantial GPU memory requirements, the generated holograms in the aforementioned approaches are primarily limited on a definition of 1080p.

Simultaneously, there have been significant efforts to generate high-definition holograms. Wu et al. \cite{holoencoder} proposed an auto-encoder architecture specifically designed for high-definition hologram generation. Similarly, Zhong et al. \cite{complex} utilized lightweight CCNNs to achieve the same objective. Both methods achieved a GPU memory usage reduction by pruning the number of network parameters. However, it is important to recognize that directly reducing network parameters may result in a limited parameter space, potentially hindering the high-quality hologram generation.

\subsection{Image Super-resolution}

Over the past few decades, numerous SR methods have been proposed, which can be categorized into four main groups: interpolation-based algorithms, reconstruction-based algorithms, CNN-based algorithms, and vision transformer (ViT)-based algorithms \cite{LEPCHA2023230}.

Interpolation-based algorithms estimate the intensity on the up-sampled grid using fixed kernels with local variance coefficients \cite{1163711} or adaptive structure kernels \cite{951537, 1658087}. However, due to their simplicity, these methods often do not provide sufficient performance to serve as holographic SR networks. Reconstruction-based algorithms assume a prior degradation model and aim to find the inverse model. Khattab et al. \cite{KHATTAB2020755} categorized these approaches into three groups: stochastic \cite{kim2013regularization, zhang2012super, shao2015posterior}, deterministic \cite{zeng2013robust, bahy2014adaptive, yadav2014better, yang2015multi}, and hybrid \cite{faramarzi2013unified, zhao2016generalized, kohler2016robust} methods. However, due to their reliance on prior image models and the solution of inverse problems, it is challenging to adapt these methods to holographic applications. Furthermore, ViT-based algorithms utilize transformers as SR networks but are not compatible with our overall framework.

On the other hand, CNN-based algorithms offer a promising choice for our task due to their superior performance and the ability to be jointly optimized. However, several challenges need to be addressed. Firstly, most SR networks focus on amplitude image SR, which may lead to a performance decline due to domain mismatch. Secondly, we need to design a lightweight SR network to accommodate the limitations of GPU memory.

\section{Proposed Method}

\begin{figure*}[htb]
 \centering 
 \includegraphics[width=0.8\linewidth]{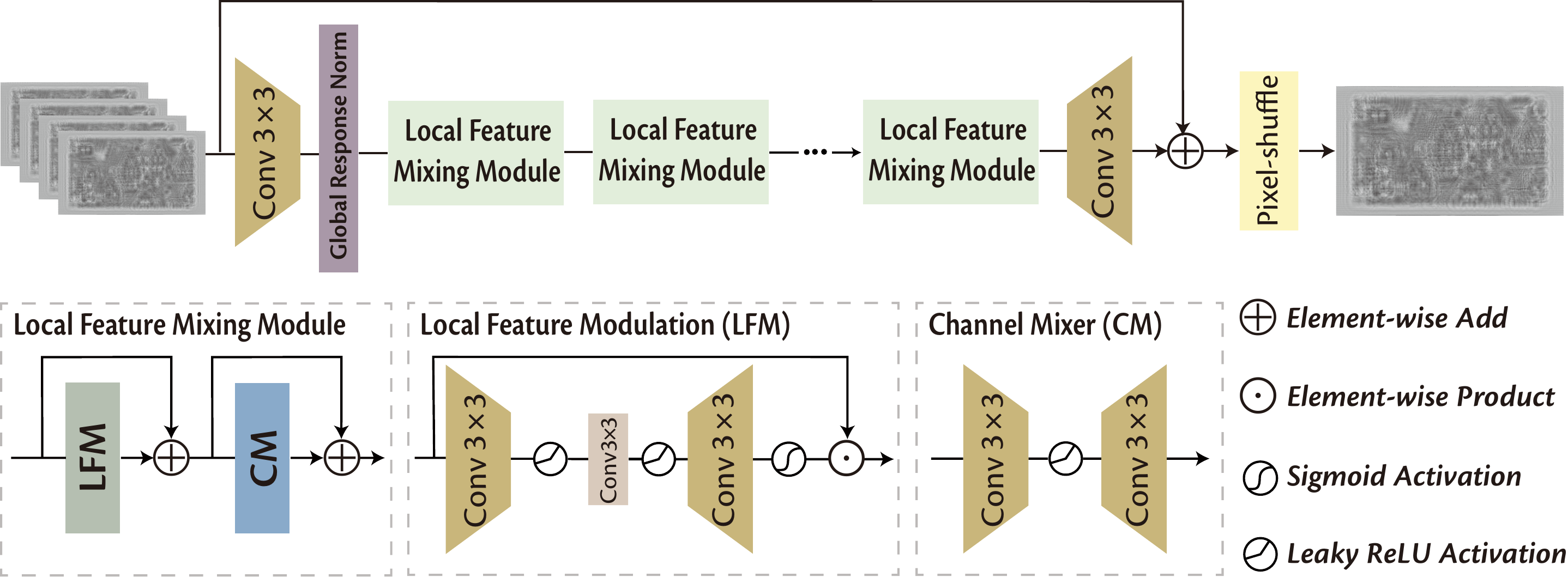}
 \caption{An overview of our proposed light-weighted holographic SR network, namely local feature mixing network (LFMN). The core module of LFMN is the local feature mixing module (LFMM) comprising a local feature modulation (LFM) module and a modified convolutional channel mixer (CCM) \cite{SAFMN} enhanced for local feature extraction.}
 \label{fig:LFMN}
\end{figure*}

We proposed a novel CGH generation framework to synthesize high-quality and ultra-high-resolution holograms, as shown in \autoref{fig:overall_architecture}. The framework consists of two main components, the phase generator and the phase encoder. For the phase generator, we employ a dividing strategy where the image is initially divided into $r^2$ sub-images with reduced resolution. These lower-definition sub-images are then fed into the phase generator network to predict the corresponding phases at the target plane. The phase and amplitude are upsampled and merged to synthesize a complex-valued wave field, which is subsequently propagated to the SLM plane using the angular spectrum method (ASM) \cite{Matsushima:09}. For the phase encoder, similar to the operations of the phase generator, we continue to apply the divide-and-conquer strategy to generate a phase-only hologram. Finally, the phase-only hologram is propagated back to the target plane, and the parameters of the networks are updated by calculating the loss between the ground truth (GT) and the reconstructed image. The formulation of each component of our method is as follows:

\begin{equation}
I_p = \mathcal{U}_{\omega}(\mathcal{O}_{\gamma}(\mathcal{D}_{\theta}(I)))   
\end{equation}
where $I_p$ is the predicted phase image, $I$ is the input image of each part, $\mathcal{U}_{\omega}(\cdot)$ is the upsampling layer, $\mathcal{D}_{\theta}(\cdot)$ is the downsampling layer, and $\mathcal{O}_{\gamma}(\cdot)$ represents the original subnetwork of CGH neural network. By this downsampling and upsampling procedure, GPU memory usage can be saved significantly. 

We will introduce the specific operations of the phase generator part and the phase encoder part in \autoref{sec:generator} and \autoref{sec:encoder}, respectively. Our proposed lightweight holographic SR network will be introduced in \autoref{sec:LFMN}. To further improve the image quality, we elaborate a pyramid architecture to synthesize a large-scale phase-only hologram in \autoref{sec:Recursive form}. The ablation study of our module will be shown in \autoref{sec:ablation}.

\subsection{Phase Generator}
\label{sec:generator}

In our approach, we apply a pixel-unshuffle layer to the input amplitude image of size $H\times W$, splitting it into $r^2$ sub-images of size $H/r \times W/r$. This step is important to avoid information loss, as demonstrated by Gu et al. \cite{gu2019self}. Here, $H\times W$ represents the dimensions of the input amplitude image. It is worth noting that the pixel-shuffle layer is only one of several potential approaches. For instance, a learnable layer or a CNN could also be considered as viable alternatives to fulfill the same objective. Then, these sub-images are fed into the phase generator network to predict $r^2$ corresponding phase patterns. To recover the full resolution from sub-phase images, we perform a pixel-shuffle layer to obtain the phase image at the original definition. Finally, the phase is combined with the input amplitude image to synthesize a complex-valued hologram at the target plane. The complex-valued hologram at full definition guarantees the retention of intricate details and enhances accuracy throughout the ASM propagation process.

\subsection{Phase Encoder}
\label{sec:encoder}

For the phase encoder part, a pixel-unshuffle layer is also employed to prevent information loss in the downsampling step. The difference between the phase generator and the phase encoder lies in the upsampling step. In contrast to the task of natural image SR, the process of hologram generation entails the transformation from the image domain (spatial domain) to the hologram domain (Fourier domain)\cite{Dong2023}. In other words, holograms and the target images do not have a direct pixel-level correspondence. Therefore, it would lead to severe degradation in the quality of the reconstructed images if a simple pixel shuffle layer is employed. To ensure the quality of generated holograms, we further design a lightweight SR network that incorporates the information between neighboring pixels in sub-holograms to generate a high-definition hologram. Similarly, the number of input channels and output channels of the phase encoder is also adjusted to $r^2$ in order to match the number of sub-images.


\begin{figure}[htb]
 \centering 
 \includegraphics[width=\linewidth]{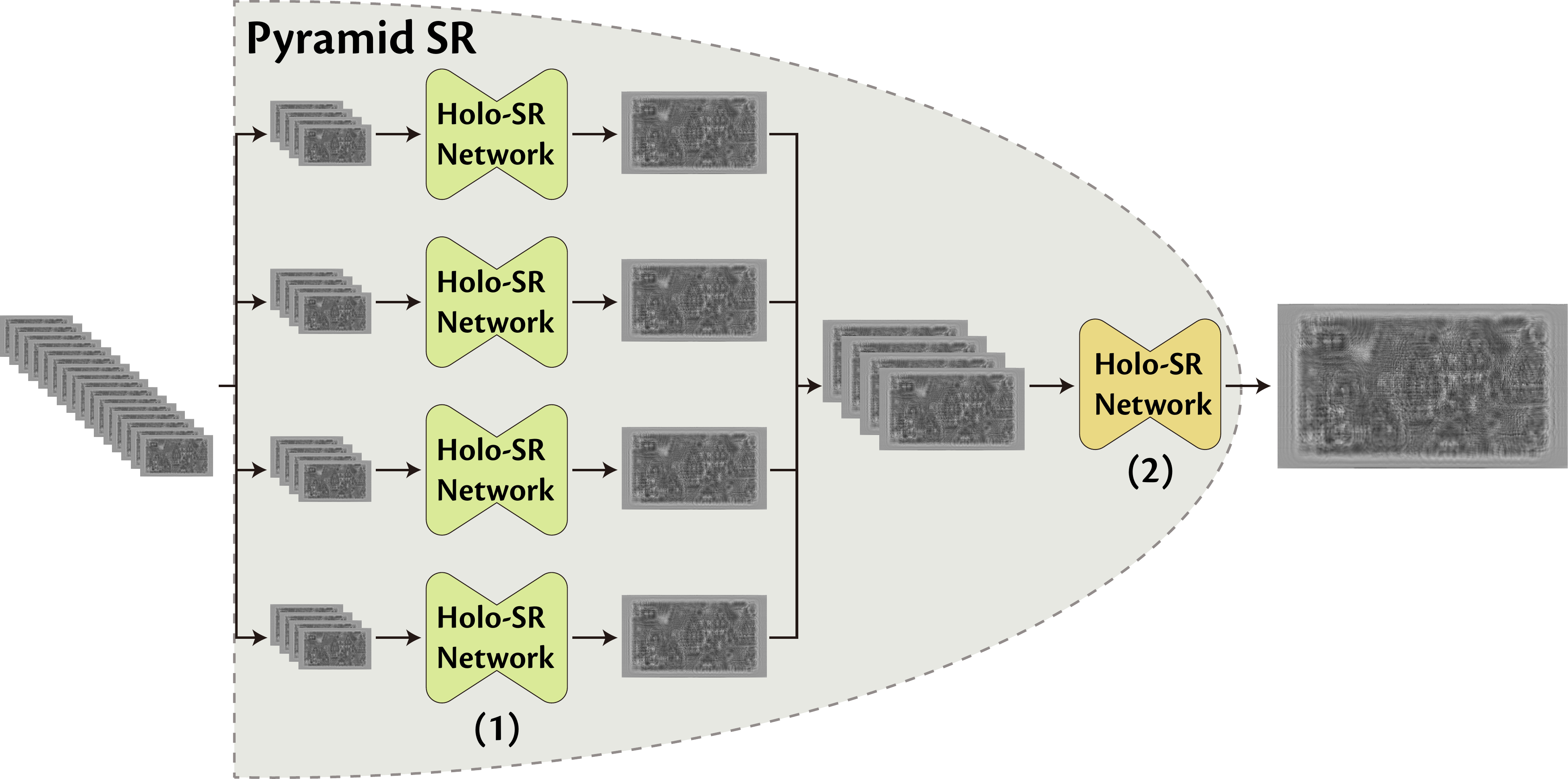}
 \caption{\textbf{Pyramid SR Network.} We design a two-stage SR architecture to synthesize a large-scale CGH.}
 \label{fig:pyramid}
\end{figure}
\subsection{Lightweight Holographic SR Network}
\label{sec:LFMN}

As illustrated in \autoref{fig:LFMN}, LFMN first transforms the $r^2$ phase-only holograms to feature maps. Then a global response normalization (GRN) \cite{GRN} is performed to normalize the feature map for better inter-channel mixing and robustness of the model. Subsequently, the normalized feature map is fed into a cascade of local feature mixing modules (LFMMs) for the extraction and mixing of local features. Finally, a pixel-shuffle operation is performed on these refined $r^2$ sub-holograms to get a single hologram at high definition. The overall architecture of our network can be formulated as follows:
\begin{equation}
\begin{aligned}
 F &= \mathcal{G}_{\alpha,\beta}(\mathcal{C}_\omega(I_l))\\
I_h &= \mathcal{PS}(\mathcal{C}_\epsilon(\mathcal{L}_\theta(F))+I_l)    
\end{aligned}
\end{equation}

\begin{figure*}[htb]
 \centering 
 \includegraphics[width=1\linewidth]{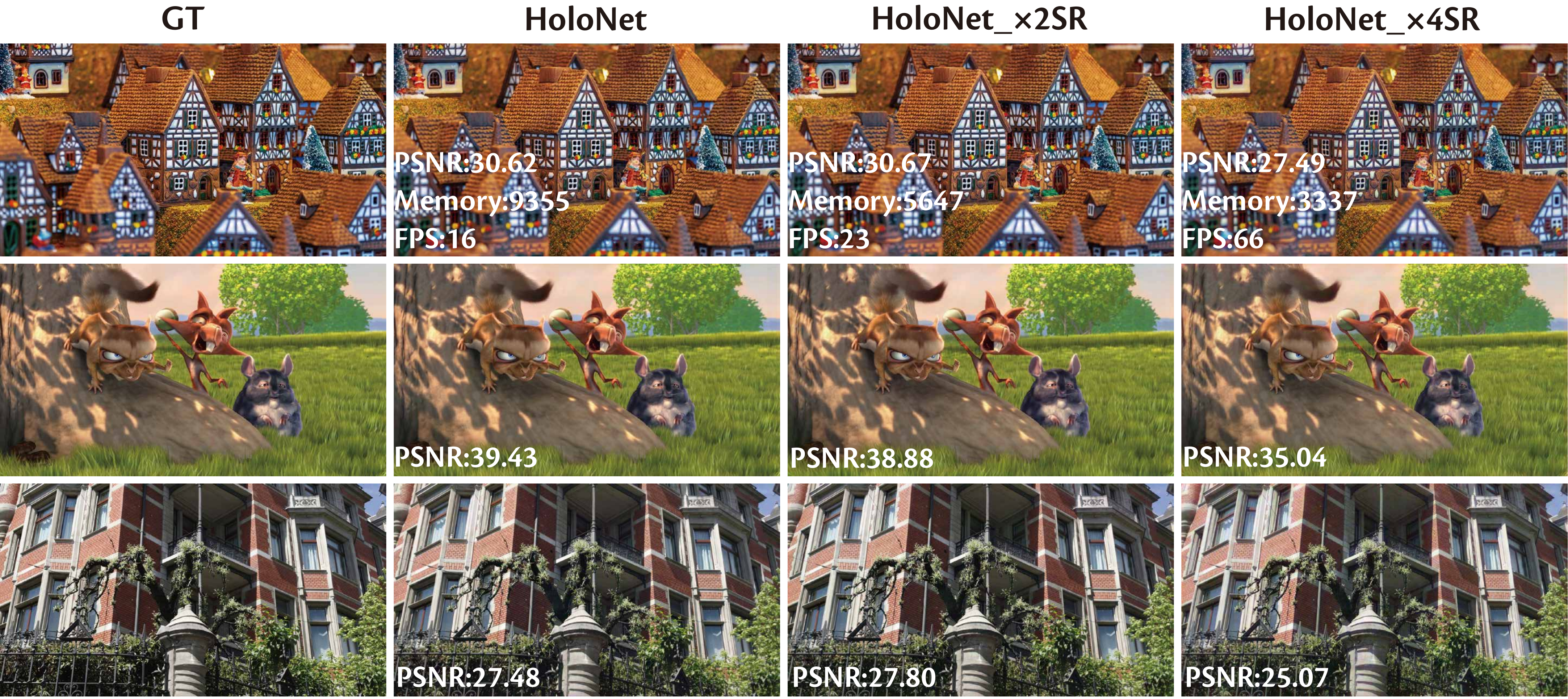}
 \caption{\textbf{Numerical reconstruction}. Full-color numerical simulations of holograms at 1080p definition generated by different methods. The second image comes from \href{www.bigbuckbunny.org}{www.bigbuckbunny.org}  (© 2008, Blender Foundation) under the Creative Commons Attribution 3.0 license (\href{https://creativecommons.org/licenses/by/3.0/}{https://creativecommons.org/licenses/by/3.0/}). The third image comes from \cite{kim2013scene}.}
 \label{fig:1080}
\end{figure*}

\begin{figure}[htb]
 \centering 
 \includegraphics[width=\linewidth]{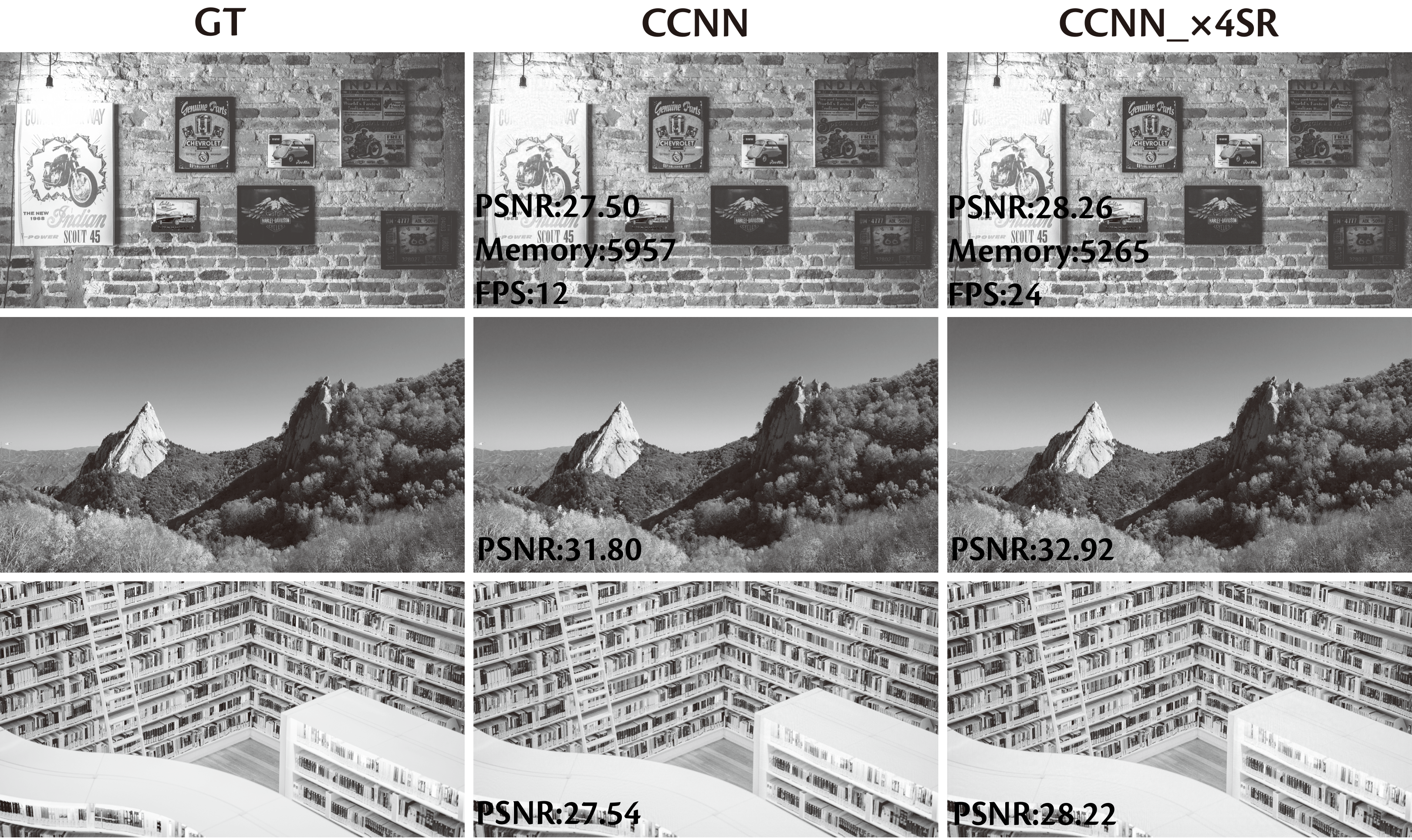}
 \caption{Numerical simulations of holograms at 4K definition generated by different methods. These images come from UHD8K \cite{UHD8K}.}
 \label{fig:4K}
\end{figure}

\noindent where $I_l$ are low-definition sub-holograms, $I_h$ is the high-definition hologram, $\mathcal{C}_\omega$ is the convolution parameterized by $\omega$ that transforms sub-holograms to feature maps, $\mathcal{G}_{\alpha,\beta}$ is GRN parameterized by $\alpha$ and $\beta$, $\mathcal{L}_\theta$ is LFMMs parameterized by $\theta$, $\mathcal{C}_\epsilon$ is the convolution parameterized by $\epsilon$ that converts feature maps back to the image space, and $\mathcal{PS}$ is the pixel-shuffle layer.

Specifically, the LFMM block comprises the LFM and enhanced CCM components, which draw inspiration from the spatial-adaptive feature modulation network (SAFMN) \cite{SAFMN}. These components have undergone significant enhancements and modifications to improve the local feature extraction and mixing capabilities within our framework. A LFMM block can be formulated as follows:
\begin{equation}
\begin{aligned}
        F' &= \text{LFM}(F)+F\\
        F'' &= \text{ECCM}(F')+F'
\end{aligned}
\end{equation}

\subsubsection{Local Feature Modulation}

In this module, inspired by the SAFM module \cite{SAFMN}, we adopt the element-wise product to modulate the feature map. However, to improve the extraction of local features, we introduce a new approach to generate the attention map.

To begin with, we generate an attention map by extracting local features from the feature map. This attention map is subsequently employed to modulate the feature map using an element-wise product. Specifically, the channels of the feature map are initially reduced by half using $3\times3$ convolutions. Then, the halved channels undergo mixing through $3\times3$ convolutions. Subsequently, the channels are restored to their original number using $3\times3$ convolutions. The attention map is activated using a Sigmoid activation function. Moreover, the activation functions between hidden layers employ the LeakyReLU activation with a slope of 0.1.

\subsubsection{Enhanced Convolutional Channel Mixer}

In this module, we further strengthen the mixing of channels by using an enhanced CCM. We modify the CCM module to achieve a better extraction of local features and less GPU memory usage. Specifically, we modify the activation functions to the LeakyReLU with the slope of 0.1 and just increase the number of channels by 1.25 for less GPU memory usage.  Furthermore, we recompose the second convolution from $1\times1$ to $3\times3$ which allows for better extraction of spatially local contexts. 

\begin{table}[htb]
  \caption{\textbf{Comparative Analysis: HoloNet vs. HoloNet w/ ours.} The comparison on the memory usage during training, inference speed and the quality of generated image. Mem represents GPU memory usage during training, measured in MiB. FPS represents the inference speed, measured in Frames Per Second (FPS). The image quality is evaluated using the metrics PSNR and SSIM. \textbf{Bolded metric} denotes the best performance. GPU memory usage is measured by nvitop \cite{nvitop}.}
  \label{tab:psnr1}
  \scriptsize%
	\centering%
\begin{adjustbox}{center}
    \scalebox{0.85}{
      \begin{tabular}{%
    c|ccc|ccc
    	}
      \toprule[1.2pt]
      \multirow{3}{*}[-1.5ex]{\textbf{Methods}}&\multicolumn{6}{c}{\textbf{HoloNet}}\\
      \cmidrule(lr){2-7}
       &\multicolumn{3}{c}{\textbf{1080p}}&\multicolumn{3}{c}{\textbf{4K}}\\
       \cmidrule(lr){2-4} \cmidrule(lr){5-7}  \noalign{\smallskip}
       &Mem$\downarrow$&FPS$\uparrow$&PSNR/SSIM$\uparrow$&Mem$\downarrow$&FPS$\uparrow$&PSNR/SSIM$\uparrow$\\
      \midrule[1pt]
      w/o &9355&16&30.47/0.93&\multicolumn{3}{c}{Out-of-memory}\\
      w/ x2&5647&23&\textbf{30.83/0.93}&14455&8&\textbf{40.29/0.96}\\
      w/ x4&\textbf{3337}&\textbf{66}&29.04/0.91&\textbf{7299}&\textbf{20}&38.05/0.98\\
      w/ pyramid&5351&20&30.51/0.93&12083&8&39.41/0.95\\
      
      \bottomrule[1.2pt]
      \end{tabular}}
\end{adjustbox}
\end{table}

\begin{table}[htb]
  \caption{\textbf{Comparative Analysis: CCNNs vs. CCNNs w/ ours.}}
  \label{tab:psnr2}
  \scriptsize%
	\centering%
	\begin{adjustbox}{center}
  \scalebox{0.85}{
    \begin{tabular}{%
    c|ccc|ccc
    	}
      \toprule[1.2pt]
      \multirow{3}{*}[-1.5ex]{\textbf{Methods}}&\multicolumn{6}{c}{\textbf{CCNNs}}\\
      \cmidrule(lr){2-7} 
       &\multicolumn{3}{c}{\textbf{4K}}&\multicolumn{3}{c}{\textbf{8K}}\\
       \cmidrule(lr){2-4} \cmidrule(lr){5-7} \noalign{\smallskip}
       &Mem$\downarrow$&FPS$\uparrow$&PSNR/SSIM$\uparrow$&Mem$\downarrow$&FPS$\uparrow$&PSNR/SSIM$\uparrow$\\
      \midrule[1pt]
      w/o &5957&12&42.06/0.97&19573&4&\textbf{46.41/0.99}\\
      w/ x4&\textbf{5265}&\textbf{24}&\textbf{42.94/0.98}&\textbf{17041}&\textbf{6}&46.31/0.98\\
      \bottomrule[1.2pt]
      \end{tabular}}
\end{adjustbox}

\end{table}

\subsection{Recursive form}
\label{sec:Recursive form}
In this section, to demonstrate that our method, like other divide-and-conquer algorithms, can be implemented in a recursive form, we design a pyramid framework to synthesize large-scale high-quality holograms to illustrate this point. Since both the pixel-shuffle layer and pixel-unshuffle layer are explicit, we only provide a pyramid SR pipeline of how to recursively merge a high-definition hologram from its sub-holograms, as illustrated in \autoref{fig:pyramid}. The complete pyramid structure can be found in the supplementary materials. 

Specifically, taking the $\times$4 pixel unshuffle as an example, in stage 1, we divide 16 sub-holograms equally into four groups, and each group passes through a lightweight SR network to obtain four $\times$2 upsampled holograms. Here we assume that the upsampling fusion process is identical for each group, allowing the network parameters of the four SR networks to be shared. Then, in stage 2, we combine the four new holograms from stage 1 again through an SR network to synthesize a large-scale hologram. It is important to note that the SR networks used in different stages are independent of each other.

\section{Results and Analysis}

\subsection{Implementation Details}

\paragraph{\textbf{Datasets}}

Following previous works \cite{Peng2020,complex}, we adopted the DIV2K dataset \cite{DIV2K} as the training data and utilized the DIV2K-val dataset as the test data for holograms at 1080p definition. However, for higher resolutions such as 4K and 8K, we randomly selected 800 images from the UHD8K training set \cite{UHD8K} as our training data and chose 100 images from the UHD8K test set as our test data.

\paragraph{\textbf{Evaluation metrics}}
To assess the quality of the reconstructed images, we employed widely used evaluation metrics, peak signal-to-noise ratio (PSNR) and structural similarity index (SSIM) \cite{1284395}. These metrics provide quantitative measures to evaluate the fidelity and similarity of the reconstructed images compared to the GT.  Furthermore, we analyzed GPU memory usage in real-time during the training process using nvitop \cite{nvitop} and compared the inference time of different methods for generating holograms.

\paragraph{\textbf{Training Details}}

The proposed framework was implemented using Python 3.8.0, and PyTorch 1.8.0. During the training, the data argumentation was performed on the input input images with random horizontal and vertical flips. We used the Adam \cite{kingma2014adam} optimizer with $\beta_1=0.9$ and $\beta_2=0.999$ to update the parameter. For our method, we set the learning rate to 5e-4. For holograms at 1080p definition, we used a combination of $l_2$ and perceptual loss \cite{johnson2016perceptual} to train our method. For holograms at 4K and 8K definitions, we used a MSE loss to train our method. All simulations were trained and inferred on an NVIDIA GeForce RTX 3090 GPU.

\paragraph{\textbf{Modification of Phase Generator and Encoder}}

Our framework modified the architecture of the phase generator and phase encoder to accommodate the division of the input image into $r^2$ sub-images. To ensure consistency, we adjusted the input channel of both components accordingly and maintained the highest number of channels in the phase generator and phase encoder unchanged. The detailed information could be found in the supplementary materials.

\begin{figure}[htb]
 \centering 
 \includegraphics[width=0.95\linewidth]{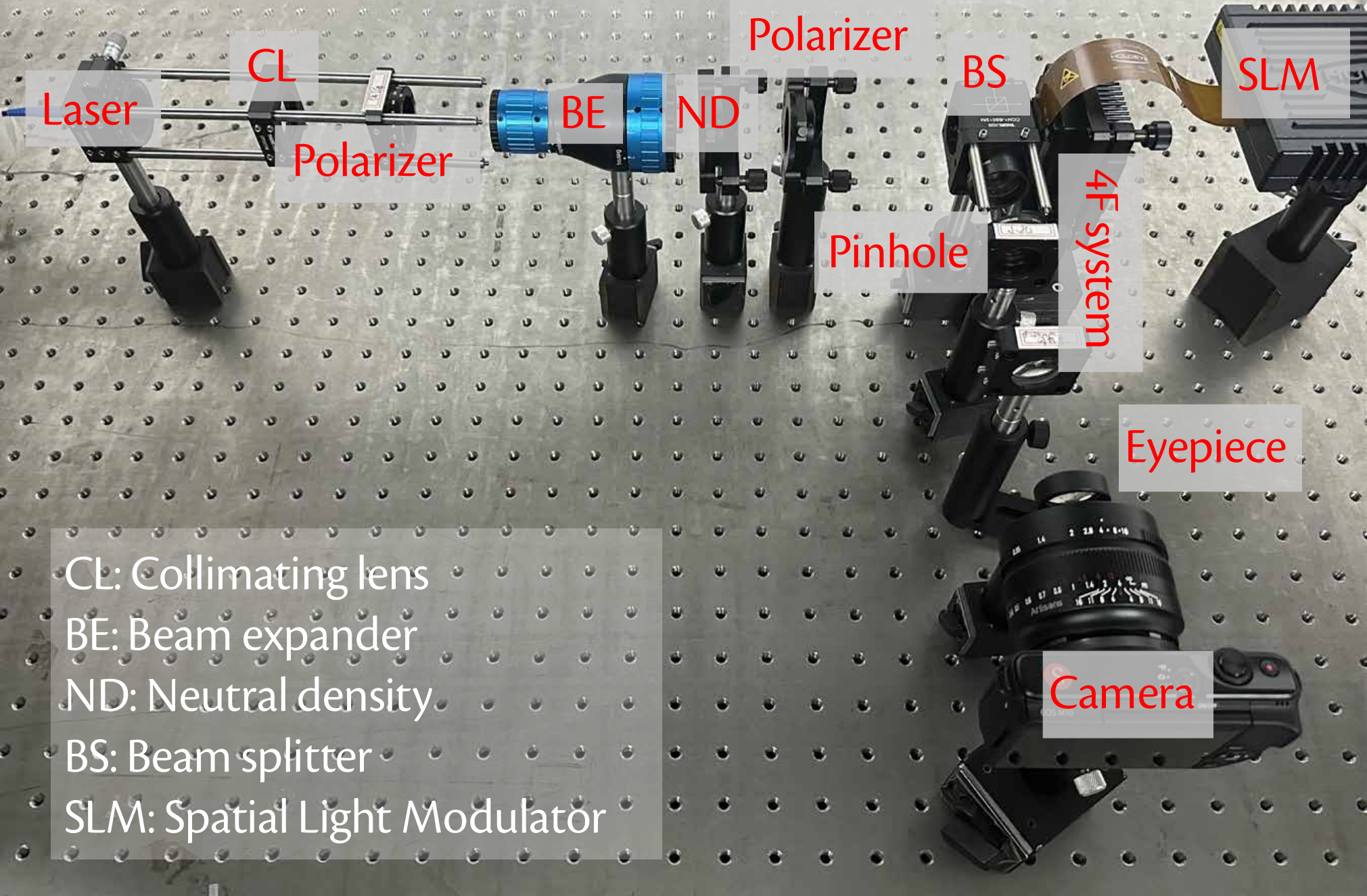}
 \caption{\textbf{Holographic display setup}. The laser produces a coherent wave field that is collimated by a collimating lens (CL). The beam expander (BE) is used to ensure that the laser incident on the SLM is as uniform as possible. Using a beam splitter (BS) cube, the  field is directed to the reflective spatial light modulator (SLM). All the experimental reconstruction images were captured using a Canon EOS M10 camera.}
 \label{fig:setup}
 \vspace{-10pt}
\end{figure}

\paragraph{\textbf{Optical Setup}}

To further validate the effectiveness of our method, we conducted VR and AR experiments in the real world. The overall optical VR setup is shown in \autoref{fig:setup}. The wavelengths of the color laser sources are 680 nm, 520 nm, and 450 nm, respectively. Specifically, to avoid color deviation, the emission intensities of the red, green, and blue channels of the multi-wavelength laser diodes are set to 100\%, 5\%, and 5\% of their respective maximum values of 40 mW in our optical setup. It should be noted that in our implementation, the color images are captured by individually exposing each wavelength and then merged in post-processing to obtain the final color image. The SLM is a HOLOEYE GAEA-2-VIS-036 with a resolution of 3840$\times$2160 and a pixel pitch of 3.74 µm. This SLM has a refresh rate of 60 Hz (monochrome) and an 8-bit depth. The collimating lens (CL) is an achromatic doublet with a focal length of 100 mm. The beam expander (BE) is used to ensure that the laser incident on the SLM is as uniform as possible. The neutral density (ND) filter is used to control the intensity of the laser, while polarization filters are used to match the polarization direction of the SLM. We provide a 4F system, where the first achromatic lens has a focal length of 60 mm, and the second achromatic lens has a focal length of 50 mm. An iris is positioned at the Fourier plane to block excessive light diffracted by the grating structure and higher-order diffractions. The focal length of the eyepiece is 50 mm. In this paper, all the experimental reconstruction images were captured using a Canon EOS M10 camera. Moreover, in the case of AR experiments, a key distinction from the VR setup is the utilization of a beam splitter to optically route holographic images, enabling their superposition onto the physical scene.

\begin{figure*}[htb]
 \centering %
 \includegraphics[width=1\linewidth]{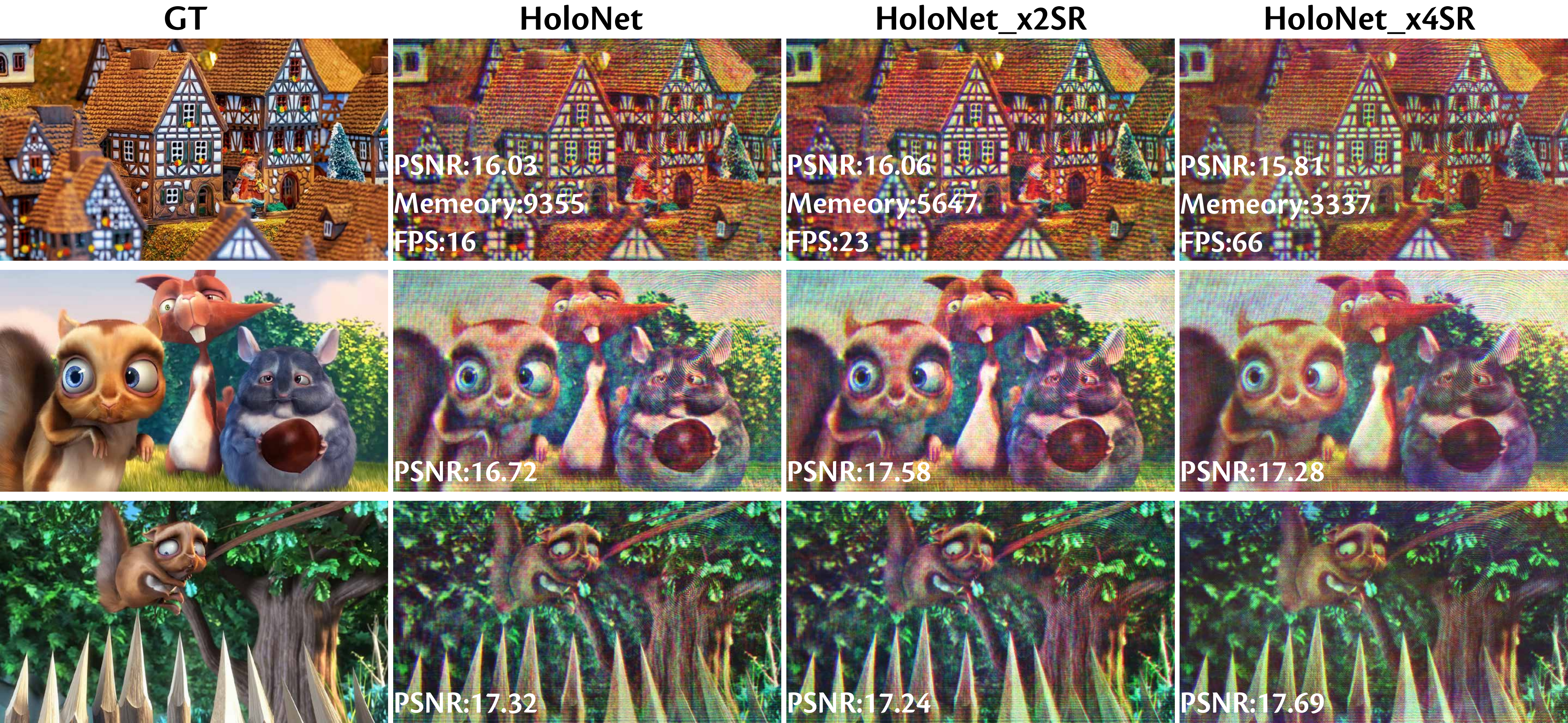}
 \caption{\textbf{Optical VR experiments}. Full-color holographic displays at 1080p resolution of holograms generated by different methods. The second and third images come from \href{www.bigbuckbunny.org}{www.bigbuckbunny.org}  (© 2008, Blender Foundation) under the Creative Commons Attribution 3.0 license (\href{https://creativecommons.org/licenses/by/3.0/}{https://creativecommons.org/licenses/by/3.0/}).}
 \label{fig:experiments}
\end{figure*}

\subsection{Comparisons against state-of-the-art}
\subsubsection{Numerical Simulations}
In this section, we validated the effectiveness of our method by integrating it into two SOTA methods, HoloNet and CCNNs, in numerical simulations. We performed simulations at a resolution of 1080p for HoloNet, considering scale factors of 2 and 4. For CCNNs, we conducted simulations at a scale factor of 4 for both 4K and 8K resolutions. 

The quantitative results are shown in \autoref{tab:psnr1} and \autoref{tab:psnr2}. For HoloNet with the proposed method, we successfully trained and inferred the 4K holograms, while naïve HoloNet was unable to achieve this due to insufficient GPU memory. As to the inference time, HoloNet with the proposed method achieved a generation rate of 66 FPS for 1080p holograms, while that of HoloNet was 16 FPS. It was worth noting that our method could be executed in real time. Moreover, HoloNet with our framework yielded a remarkable reduction in memory footprint by 64.3\%. The PSNR of reconstructed images was measured at 29.04 dB during the simulation, which was 1.43 dB slightly lower than that by the baseline HoloNet. To further improve the image quality, in the pyramid architecture, we achieved a PSNR of 30.51 dB for image reconstruction, which was 0.04 dB higher than the baseline HoloNet. At the same time, our method also reduced the required GPU memory by 42.8\%. Similar trends were observed in the CCNNs with the proposed method. By integrating our method into CCNNs, we were able to reduce GPU memory by 12.9\% while achieving a 0.88 dB higher PSNR for image reconstruction compared to CCNNs in the 4K definition. The inference time of CCNNs with our method was 24 FPS, while the inference time of CCNNs was only 12 FPS. Particularly, we successfully trained 8K definition holograms on a consumer-grade GPU card for the first time in simulations. The 8K definition holograms and reconstructed images can be found in the supplementary materials. 

Furthermore, we compared the visual quality with the SOTA method, as shown in \autoref{fig:1080} and \autoref{fig:4K}. We observed our method could achieve holographic image quality that matched or even surpassed SOTA methods, despite of the significantly reduced GPU memory usage and accelerated hologram inference speed. These results unequivocally demonstrated that our proposed framework facilitates more efficient utilization of available resources without sacrificing performance.

\subsubsection{Optical Experiments}
In addition to the evaluation of simulated results, we conducted VR experiments at 1080p resolution in scenes containing significant texture details, as illustrated in \autoref{fig:experiments}. Deep learning-based CGH methods can generate high-fidelity display images without speckle noise because of the smooth phase distribution in the reconstructed plane \cite{Dong2023}. The observations clearly indicated that the holograms generated by our method achieved display quality comparable to that of HoloNet.Furthermore, our method demonstrated significant advantages in terms of GPU memory usage and hologram generation time. By employing a divide-conquer-and-merge strategy, we were able to reduce memory usage by up to 64.3\% and improve inference speed by up to 3 $\times$ compared to baseline HoloNet.  These improvements allowed us to successfully train 4K holograms, as well as generate real-time and high-quality 1080p holograms on consumer-grade GPUs. More information can be found in the supplementary video.

During hardware capture, we encountered challenges such as ringing artifacts at the edges and color non-uniformity. These issues hindered the reconstruction process from achieving results on par with the simulated outcomes. It is important to note that these challenges stem from the non-ideal nature of the hardware prototype itself and are unrelated to the framework. To address these limitations, we intend to explore and implement CITL optimization strategies in future work. By leveraging these strategies, we aim to enhance the quality of our display and mitigate the observed hardware-related issues. However, these are beyond the scope of this work.

\begin{figure}[htb]
 \centering 
 \includegraphics[width=\linewidth]{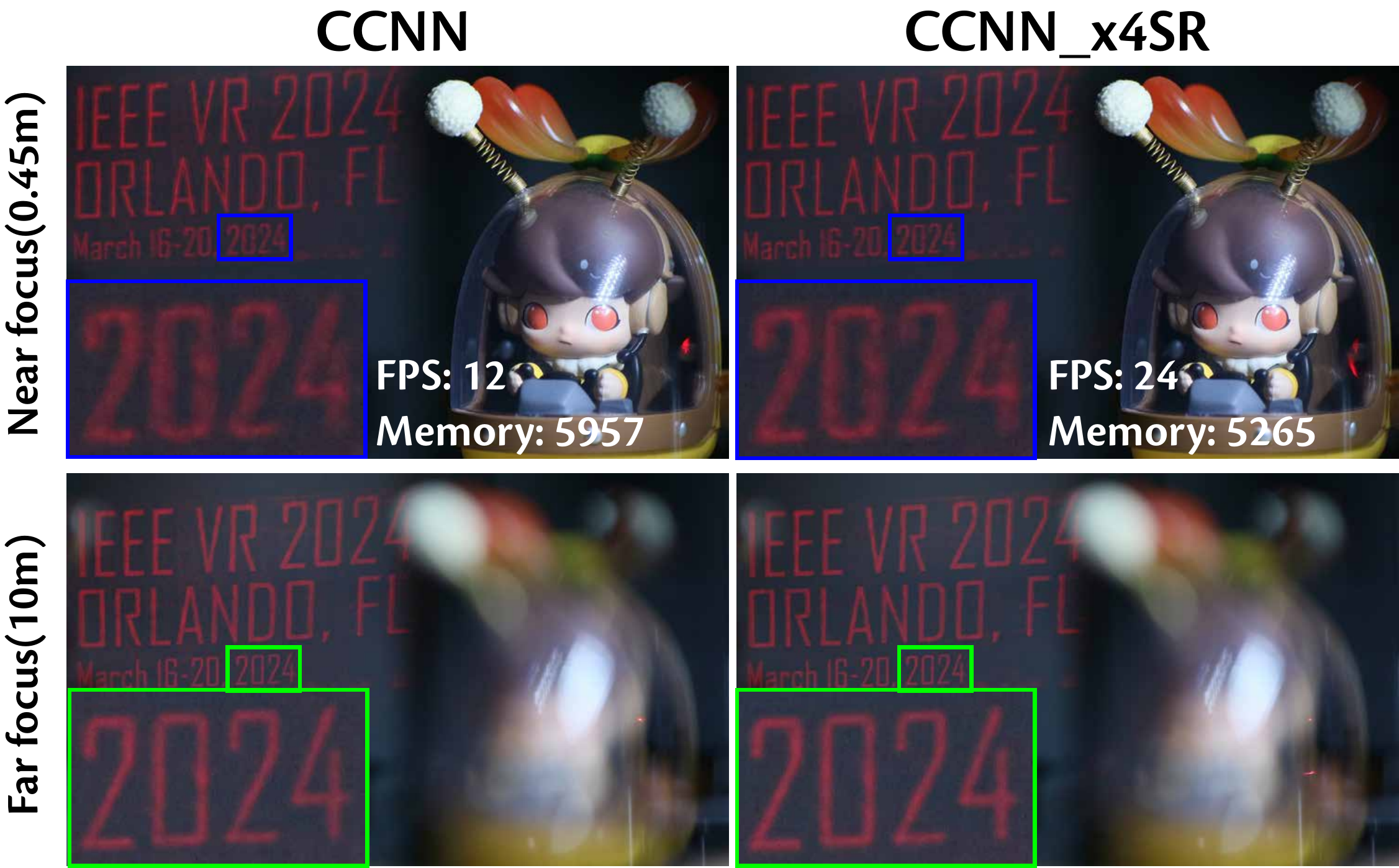}
 \caption{\textbf{Optical AR experiments}. An AR experiment at 4K resolution. Specifically, we placed a Pop Mart toy at a distance of 0.45m and projected the holographic display content through an eyepiece at a distance of 10m.}
 \label{fig:AR}
\end{figure}

To further demonstrate the effectiveness of the proposed method, we conducted the AR experiment at 4K resolution in the real world, as shown in \autoref{fig:AR}. Specifically, we placed a Pop-Mart toy at a distance of 0.45m and projected the holographic display content through an eyepiece at a distance of 10m. By adjusting the camera parameters, we were able to observe the focusing and defocusing effects of both real and virtual content at different distances. We can clearly see that our method achieves display quality that is on par with CCNNs, providing further evidence of the effectiveness of our method.

\section{Ablation Study}
\label{sec:ablation}

In this section, we further conducted ablation studies. The ablation experiments were conducted under red light, and holograms were at 1080p definition. The training set was DIV2K, and the results of \autoref{tab:ablation} was measured under DIV2K-val.

\begin{table}[htb]
  \caption{\textbf{Ablation Study.} We conducted ablation experiments to validate the effectiveness of our proposed framework and SR network LFMN. The definition of holograms was 1080p and the light color was red in experiments. ECCM$\rightarrow$CCM denotes that change the strengthened CCM to the original CCM. ASM at low definition denotes that the complex-valued sub-holograms generated by the phase generator are directly propagated respectively at low definition without upsampling by the pixel-shuffle layer.}
  
  \label{tab:ablation}
  \scriptsize%
	\centering%
  \begin{tabu}{%
	*{5}{c}%
	}
  \toprule
  Ablation (1080p)&Variant&Scale&PSNR/SSIM\\
  \midrule
  \midrule
  HoloNet&/&/&30.47/0.9257\\
  \midrule
   \multirow{2}{*}{Proposed}&/&$\times$2&\textbf{30.83/0.9312}\\
   &/&$\times$4&29.04/0.9139\\
   \midrule
   \multirow{4}{*}{Framework}&ASM at low definition&$\times$2&16.63/0.4893\\
   &SR$\rightarrow$None&$\times$2&29.83/0.9165\\
   &SR$\rightarrow$None&$\times$4&28.55/0.9007\\
   \midrule
   \multirow{5}{*}{LFMN}&LFM$\rightarrow$None&$\times$2&30.08/0.9147\\
   &ECCM$\rightarrow$None&$\times$2&30.35/0.9167\\
   &GRN$\rightarrow$None&$\times$2&30.13/0.9085\\
   &ECCM$\rightarrow$CCM&$\times$2&30.73/0.9244\\
   &LFM$\rightarrow$SAFM&$\times$2&30.21/0.9141\\
  \bottomrule
  \end{tabu}%
\end{table}

\subsection{ASM Propagation at Full Definition}

To verify the effectiveness of ASM propagation at full definition, we propagated sub-holograms predicted by the phase generator respectively. It was important to note that during the propagation process, the original size of the image was maintained by increasing the pixel pitch of the SLM to $p/r$, where $p$ represents the SLM pixel pitch. We found that when we propagated sub-holograms at low definition respectively, there existed a significant decline in performance. The results demonstrated the significance of propagating with high accuracy.

\subsection{SR Network LFMN}

By substituting a pixel-shuffle layer for LFMN, we found a performance decline of 1.0 dB and 0.49 dB for the scale factor of 2 and 4, respectively. These results can validate the effectiveness of LFMN.

\paragraph{\textbf{GRN, enhanced CCM and LFM.}}
We observed that the removal of GRN, enhanced CCM, and LFM resulted in performance decreases of 0.7 dB, 0.48 dB, and 0.75 dB, respectively. Furthermore, when replacing the enhanced CCM and LFM with the original modules in the SAFMN \cite{SAFMN}, we found performance decreases of 0.1 dB and 0.62 dB, respectively.
These results highlighted the effectiveness of our proposed modules in holographic SR.


\begin{table}[htb]
  \caption{The memory utilization of different model components within the CCNNs integrated with our proposed method during the training of holograms at various definitions.}
  
  \label{tab:ASM}
  \scriptsize%
	\centering%
  \begin{tabu}{%
	*{4}{c}%
	}
  \toprule
  \multirow{2}{*}{Components}&\multicolumn{3}{c}{Memory(MiB)}\\
  &1080p&4K&8K\\
  \midrule
  \midrule
  ASM&396&1648&6548\\
  Holo-SR&194&786&3130\\
  CCNNs&158&642&2562\\
  \bottomrule
  \end{tabu}%
\end{table}

\section{Discussion}

In this manuscript, to accurately calculate the inference time for each network, we used \textit{torch.cuda.event} to measure time on the GPU. It is crucial here to utilize \textit{torch.cuda.synchronize()}. This line of code performs synchronization between the host and device (i.e., GPU and CPU), so the time recording takes place only after the process running on the GPU is finished. This overcomes the issue of unsynchronized execution.

\section{Limitation and Future work}

Firstly, we found that the improvements achieved by integrating our method into CCNNs were not as significant as those obtained by integrating it into HoloNet. To provide a more detailed analysis of this phenomenon, we have listed the memory usage of each module in CCNNs in \autoref{tab:ASM}. It can be observed clearly that the memory usage of ASM propagation is higher than that of the neural network, which indicates that the primary contributor to the overall memory usage is the ASM propagation process. Therefore, the effectiveness of our method in lightweight CGH networks may be diminished to some extent. In the future, we will further explore methods to reduce the memory requirements of ASM propagation to achieve higher-definition hologram generation.

Secondly, in this manuscript, we only applied the divide-conquer-and-merge strategy to the 2D CGH generation framework. Recently, there have been a few SOTA 3D holographic generation networks \cite{Shi2021, Shi2022} and holographic compression frameworks \cite{Shi:22, DONG2023102464}. Therefore, in the future, we will further integrate our method into more holographic architectures to demonstrate the generalizability of our approach and achieve real-time generation and transmission of ultra-high-resolution holograms.

\section{Conclusion}
In conclusion, we proposed a divide-conquer-and-merge strategy to address the memory and computational capacity scarcity in large-scale CGH generation. By incorporating our proposed method into current SOTA CGH neural networks, we achieved significant reductions in GPU memory usage and improvements in the inference time without compromising the image quality. In particular, we successfully trained and inferred 8K definition holograms on an NVIDIA GeForce RTX 3090 GPU for the first time. We believe that our framework can provide a new way of thought for overcoming the memory barrier encountered in ultra-high-definition hologram generation. In the future, we aim to further optimize the memory usage of current CGH neural networks to generate holograms at 16K+ definition.

\acknowledgments{
The authors wish to thank Chao Xu for his invaluable assistance with the experimental system. This work was supported by National Key Research and Development Program of China (2023YFB3611501).}

\bibliographystyle{abbrv-doi}

\bibliography{template}
\end{document}